\documentclass[12pt,twoside,a4paper,amsmath,amssymb,apsmy,showkeys,showpacs,nofootinbib]{revtex4}

\usepackage[cp1251]{inputenc}
\usepackage[russian,english]{babel}
\usepackage[left=2.5cm, right=2.5cm, top=2.5cm, bottom=2.2cm, bindingoffset=0cm]{geometry}

\usepackage[T2A]{fontenc}

\usepackage{subfig}
\usepackage{graphicx}
\usepackage{graphics}
\usepackage{epsfig}
\usepackage{float}
\usepackage{multirow}
\usepackage{bigstrut}
\usepackage{latexsym}
\usepackage{array}
\usepackage{dcolumn}
\usepackage{bm}

\begin{document}
\thispagestyle{myheadings}
	
\title{Coherent radiation produced by microtron electron bunches}
\author{S.V. Anishchenko}
\email{sanishchenko@mail.ru}
\affiliation{Research Institute for Nuclear Problems of Belarusian State University \\Bobruiskaya Str. 11, 220006 Minsk, Belarus}

\author{P.V. Molchanov}
\email{molchanov-p-v@mail.ru}
\affiliation{Research Institute for Nuclear Problems of Belarusian State University \\Bobruiskaya Str. 11, 220006 Minsk, Belarus}

\begin{abstract}
The theory of coherent transition radiation produced by a relativistic electron beam during its extraction from a microtron is established. Expressions for the beam form factor, spectral-angular and angular distribution of coherent transition radiation are obtained in explicit form. Estimates of microwave noise caused by coherent transition radiation are given.
\end{abstract}

\maketitle

\section{Introduction}
One of the most important challenges of modern physics is the design of coherent electromagnetic radiation sources based on electron beam accelerators.
In this regard, an important part of research activity planned to be carried out at the LINAC-200 linear electron accelerator and the Microtron 25-MT (JINR, Dubna) within the framework of the FLAP~\cite{2021Baldin} collaboration is associated with the design of coherent electromagnetic radiation sources operating in the microwave and teraherz ranges.
The maximum electron energy in the linear accelerator, scheduled to be commissioned at the end of 2025, is more than 200 MeV, while in the microtron, which has been operating since 2013, it is up to 25 MeV.

The beams of charged particles accelerated in LINAC-200 and the microtron consist of electron bunches following one another at a frequency~$f$. This makes it possible to obtain coherent electromagnetic radiation when beams of particles propagate through various electrodynamic structures.
Typical radiation frequencies must be proportional to the harmonics of the fundamental frequency~$f$. The spectral-angular distribution of electromagnetic radiation is determined by the beam form factor and the spectral-angular distribution of spontaneous electron emission~\cite{1982VG}.

To measure the radiation properties, it is proposed to use high-frequency horn antennas located at some distance from the electrodynamic structure. In this case, a problem arises related to the correct assessment of noise. The noise should be lower than the measured signal. In an accelerator system, the source of noise can be coherent transition radiation generated by the electron beam from the vacuumized output channel.
Note that, in contrast to the incoherent spontaneous transition radiation emitted at small angles $\theta\sim1/\gamma$ ($\gamma$ --- Lorentz factor) to the electron velocity, the angular distribution of coherent transition radiation demonstrates also a noticeable increase at large angles~\cite{2003Serov}.

In this regard, the present work is devoted to the evaluation of the spectral-angular distribution of coherent transition radiation generated by the electron beam during its extraction from the microtron output channel into the environment. The paper outlines as follows. First, we briefly describe the scheme of beam extraction from the microtron. Then we obtain expressions for the electron beam form factor, angular and spectral-angular distribution of coherent transition radiation. The work concludes with estimates of the noise caused by coherent transition radiation generated by the microtron beam.

\section{Microtron electron bunches}
\begin{figure}[ht]
	\begin{center}
		\resizebox{160mm}{!}{\includegraphics{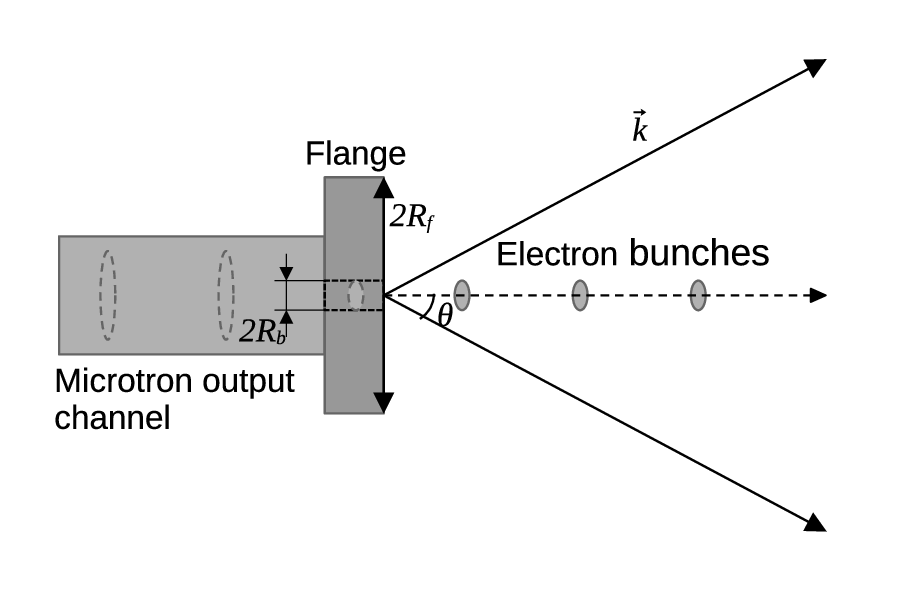}}\\
	\end{center}
	\caption{Transition radiation generated by microtron electron bunches.}	\label{fig:geometry}
\end{figure}

The electron beam is extracted from the microtron using a vacuumized output channel (see figure \ref{fig:geometry}). The average
beam current during is $I_b=10$~mA. The electric current pulse lasts for~$T_b=5.2$~$\mu$s. Each pulse is a sequence of electron
bunches following each other at~$f\approx2.8$~GHz. The length of one bunch, which has a Gaussian 
profile in the longitudinal direction, is $\sigma=1$--$3$~mm.
(For estimates, we will use the value $\sigma=2.4$~mm which is typical for other similar accelerators~\cite{2017Naumenko}.) The transverse size of the beam is determined by the radius of the hole $R_b=0.6$~cm in the microtron output channel ending with a metal flange with a radius~$R_f\approx3.5$~cm.
It should be noted here that, despite the fact that the microtron is designed for a maximum particle energy~25~MeV, the most preferable energy is one that does not exceed~10~MeV. In this case, no additional monitoring of induced radioactivity is required.

Knowing the duration of the current pulse $T_b$ and the repetition rate of electron bunches, we can estimate the bunch number in one pulse $N_b\approx fT_b=1.4\cdot10^4$. The number of particles in a single bunch~$N_e$ is determined by dividing the average current $I_b$ by the product of the frequency $f$ and the elementary charge $q_e$: $N_e= I_b/fq_e\approx2.3\cdot10^7$. We will use the obtained values of $N_b$ and $N_e$ in further estimates.

\section{Transition radiation of electron bunches}
Let an electron beam of radius $R_b$ produced by the microtron move along the $z$ axis. Then the position distribution of electrons normalized to the particle number in the beam $N_bN_e$ can be written as follows
\begin{equation}
	f(\vec r)=\frac{N_e}{\pi R_b^2}\sum_{j=1}^{j=N_b}\frac{1}{\sqrt{2\pi}\sigma}\exp\Big(-\frac{(z-j\Delta z)^2}{2\sigma^2}\Big).
\end{equation}
Here, $\Delta z=\beta c T$ denotes a distance between two bunches, which is expressed through particle velocity $\beta c$ and $T=1/f$.

The Fourier transform of the function $f(\vec r)$ allows us to find the form factor of the beam
 $|F|^2=|\int_{-\infty}^{+\infty}f(\vec r)\exp(-i\vec r\vec k)d^3\vec r|^2$: 
\small\begin{equation}
	\label{eq:ff}
	|F(\omega,\theta)|^2=N_bN_e^2\cdot\frac{\sin^2\Big(\frac{1}{2}N_b\omega T\beta\cos\theta\Big)}{N_b\sin^2\Big(\frac{1}{2}\omega T\beta\cos\theta\Big)}\cdot\exp\Big(-\frac{\omega^2\sigma^2\beta^2\cos^2\theta}{c^2}\Big)\cdot\Big(\frac{2J_1\big(\omega R_b\sin\theta/c\big)}{\omega R_b \sin\theta/c}\Big)^2.
\end{equation}
Due to the presence of a multiplier containing the ratio of sines in \eqref{eq:ff}, which is caused by the periodic structure of the beam, the form factor~$|F|^2$ is characterized by a sequence of maxima at angle-dependent  frequencies
\begin{equation}
	\label{eq:omegan}
	\omega_n(\theta)\approx\frac{n\omega_0}{\beta\cos(\theta)}
\end{equation}
and full widths at half maximum
\begin{equation}
	\Delta\omega_n(\theta)\approx\frac{2.8\omega_0}{N_b\beta\cos(\theta)}.
\end{equation}
Here, $n$ is a non-negative integer. In the case of a microtron, we have~$\Delta\omega_n/\omega_n\approx10^{-4}$. The appearance of the indicated frequency components should be expected in the spectral-angular distribution of the coherent transition radiation generated by the electron beam from the microtron. Let us note that the frequencies of coherent transition radiation at angles $\theta\approx1/\gamma$ exceed the frequencies of electric current harmonics $nf$ by $n\omega_0/2\pi\gamma^2\approx7n$~MHz at a particle energy~10~MeV. Let us also pay attention to the presence of broadband megaherz radiation corresponding to $n=0$.

\begin{figure}[ht]
	\begin{center}
		\resizebox{120mm}{!}{\includegraphics{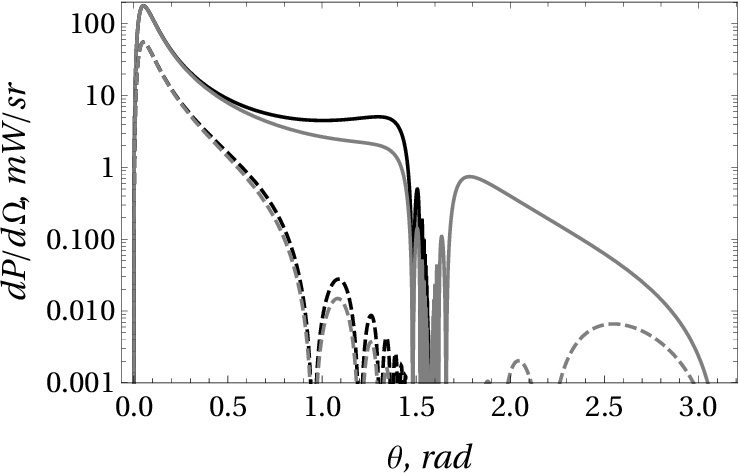}}\\
	\end{center}
	\caption{The angular distributions of coherent transition radiation corresponding to harmonics with indices $n=1$ (solid curves) and $n=8$ (dashed curves), respectively. Black and gray curves correspond to the condition $\omega_n(\theta)R_f\sin\theta/c\gtrsim\pi/2$ and $\omega_n(\theta) R_f\sin\theta/c\lesssim\pi/2$, respectively.}	\label{fig:angular}
\end{figure}

\begin{figure}[ht]
	\begin{center}
		\resizebox{120mm}{!}{\includegraphics{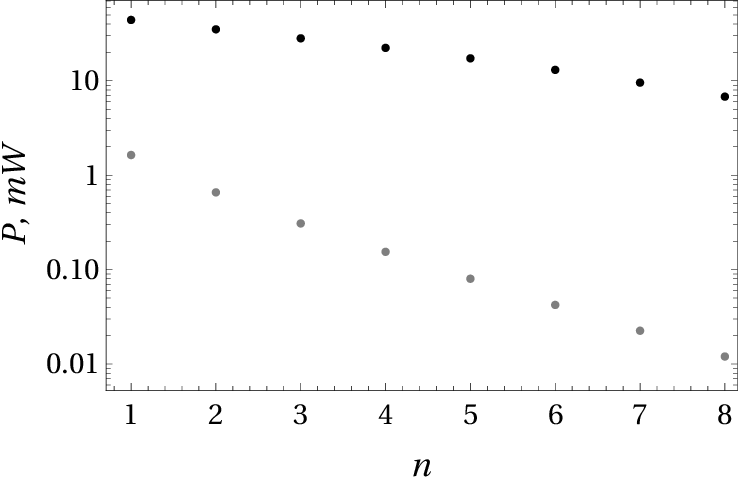}}\\
	\end{center}
	\caption{Radiation power for different harmonics. Black (gray) dots correspond to radiation emitted into the forward (backward) hemisphere in respect to the direction of particle motion.} \label{fig:power}
\end{figure}

Since the radius of the output flange $R_f$ exceeds the beam radius $R_b$ significantly, we can neglect the difference in the spectral-angular distributions of spontaneous emission $\frac{\partial^2 W_e}{\partial\Omega\partial\omega}$ emitted by different particles. This means that the spectral-angular distribution of the beam radiation can be written as a product~\cite{1982VG}
\begin{equation}
	\label{eq:dWb}
	\frac{\partial^2W}{\partial\Omega\partial\omega}=\frac{\partial^2W_e}{\partial\Omega\partial\omega}|F(\omega,\theta)|^2.
\end{equation}

According to \cite{1982VG}, the spectral-angular distribution $\frac{\partial^2 W_e}{\partial\Omega\partial\omega}$ can be written as follows
\begin{equation}
	\label{eq:dWeVG}
	\frac{\partial^2 W_e}{\partial\Omega\partial\omega}=\frac{q_e^2\omega^2}{4\pi^2c^3}\int_{t_1}^{t_2}\vec E_{-\vec k}^{(+)s}(\vec r(t),\vec v(t))e^{i\omega t}\vec E_{-\vec k}^{(+)s*}(\vec r(\tau),\vec v(\tau))e^{-i\omega \tau}dtd\tau.
\end{equation}
Here, $E_{-\vec k}^{(+)s}$ describes scattering of a plane wave with a wave vector $-\vec k$ on the microtron flange. 
In the case of a conducting plate of an infinite radius, the explicit expression for~$\frac{\partial^2 W_e}{\partial\Omega\partial\omega}$ was obtained by Ginzburg and Frank~\cite{1996Ginzburg}
\begin{equation}
	\label{eq:dWe}
	\frac{\partial^2 W_e}{\partial\Omega\partial\omega}=\frac{q_e^2}{\pi^2c}\frac{\beta^2\sin^2\theta}{(1-\beta^2\cos^2\theta)^2}.
\end{equation}
Both a relativistic particle and the alternating currents induced in the conducting plate by the particle field contribute to \eqref{eq:dWe}. Since the flange radius has finite dimensions, the radiation from the induced currents at frequencies satisfying the inequality $\omega R_f\sin\theta/c\lesssim\pi/2$ should be strongly suppressed. In fact, in this case we are dealing with radiation from a instantly accelerated charge, which corresponds to a spectral-angular distribution of the form~\cite{1963Jackson}
\begin{equation}
	\label{eq:dWef}
	\frac{\partial^2 W_e}{\partial\Omega\partial\omega}=\frac{q_e^2}{4\pi^2c}\frac{\beta^2\sin^2\theta}{(1-\beta\cos\theta)^2},
\end{equation}
Expression \eqref{eq:dWef} practically coincides with \eqref{eq:dWe} in the region of small angles $\theta\sim1/\gamma$, where the maxima of both spectral-angular distributions occur at ultrarelativistic particle energies ($\gamma\gg1$). A noticeable difference between the distributions is observed only at large angles $\theta\gtrsim\pi/2$.
Note that \eqref{eq:dWef} coincides with the distribution of electromagnetic radiation during beta decay emitted by a instantly appearing electron.

Integration over the frequencies of the spectral-angular distribution near $\omega_n$ allows us to find the contributions to the angular distribution of distinct harmonics:
\begin{equation}
	\label{eq:dWn}
	\frac{dW_n}{d\Omega}=N_bN_e^2\frac{q_e^2\omega_0}{\pi^2c}\frac{\beta\sin^2\theta}{(1-\beta^2\cos^2\theta)^2|\cos(\theta)|}\cdot\exp\Big(-\frac{n^2\omega_0^2\sigma^2}{c^2}\Big)\cdot\Big(\frac{2J_1\big(n\omega_0R_b\tg\theta/\beta c\big)}{n\omega_0 R_b\tg\theta/\beta c}\Big)^2
\end{equation}
at $\omega_n R_f\sin\theta/c\gtrsim\pi/2$ and
\begin{equation}
	\label{eq:dWn2}
	\frac{dW_n}{d\Omega}=N_bN_e^2\frac{q_e^2\omega_0}{\pi^2c}\frac{\beta\sin^2\theta}{4(1-\beta\cos\theta)^2|\cos(\theta)|}\cdot\exp\Big(-\frac{n^2\omega_0^2\sigma^2}{c^2}\Big)\cdot\Big(\frac{2J_1\big(n\omega_0R_b\tg\theta/\beta c\big)}{n\omega_0 R_b\tg\theta/\beta c}\Big)^2
\end{equation}
at $\omega_n R_f\sin\theta/c\lesssim\pi/2$.

Dividing $\frac{dW_n}{d\Omega}$ by the current pulse time $N_bT$, we obtain the angular distribution of the radiation power corresponding to the $n$-th harmonic
\begin{equation}
	\label{eq:Pne}
	\frac{dP_n}{d\Omega}=N_e^2\frac{q_e^2\omega_0^2}{2\pi^3c}\frac{\beta\sin^2\theta}{(1-\beta^2\cos^2\theta)^2|\cos(\theta)|}\cdot\exp\Big(-\frac{n^2\omega_0^2\sigma^2}{c^2}\Big)\cdot\Big(\frac{2J_1\big(n\omega_0R_b\tg\theta/\beta c\big)}{n\omega_0 R_b\tg\theta/\beta c}\Big)^2
\end{equation}
at $\omega_n R_f\sin\theta/c\gtrsim\pi/2$ and
\begin{equation}
	\label{eq:Pnf}
	\frac{dP_n}{d\Omega}=N_e^2\frac{q_e^2\omega_0^2}{2\pi^3c}\frac{\beta\sin^2\theta}{4(1-\beta\cos\theta)^2|\cos(\theta)|}\cdot\exp\Big(-\frac{n^2\omega_0^2\sigma^2}{c^2}\Big)\cdot\Big(\frac{2J_1\big(n\omega_0R_b\tg\theta/\beta c\big)}{n\omega_0 R_b\tg\theta/\beta c}\Big)^2.
\end{equation}
at $\omega_n R_f\sin\theta/c\lesssim\pi/2$.

Figure \ref{eq:ff} shows angular distributions for two different harmonics calculated using formulas~\eqref{eq:Pne} and \eqref{eq:Pnf}. The suppression of radiation with increasing $n$, which is caused by the finite length of an electron bunch~$\sigma$, is clearly visible. A distinctive feature of the angular distribution~\eqref{eq:Pnf} is the possibility of observing transition radiation in the backward hemisphere relative to the direction of particle motion.
Integrating the distribution of radiation power over angles numerically, we obtain the power values corresponding to different harmonics (figure \ref{eq:ff}). 

Analysis of the obtained values allows us to conclude that at an electron energy of 10~MeV, the 
radiation power into the forward hemisphere exceeds the radiation power into the back hemisphere 
by approximately two orders of magnitude. When shifting attention from the first harmonic to the 
eighth, the radiation power into the forward (backward) hemisphere changes from 44~mW to 7~mW 
(from 1.6~mW to 0.01~mW). The specified values, as well as the angular and spectral-angular distributions of the transition radiation, should be taken into account when conducting experiments: the measured signal should not be lower than the parasitic noise caused by the transition radiation.

\section{Conclusion}
In this paper, a theory of coherent transition radiation generated by a relativistic electron beam from a microtron is established. Expressions for the electron beam form factor, as well as the spectral-angular and angular distribution of coherent transition radiation are obtained in explicit form.
It is demonstrated that the radiation spectrum consists of a sequence of multiple harmonics. Their frequencies and FWHMs are inversely proportional to the cosine of the angle measured from the system axis. Using the developed theory, the estimates of the microwave noise created by the electron beam from a microtron are obtained.
The presence of this parasitic noise must be taken into account when measuring coherent radiation from the microtron beam passing through electrodynamic structures.

The authors thanks Professor V.G. Baryshevsky and A.A. Gurinovich for valuable comments and discussions of the results obtained, as well as the FLAP Collaboration for discussions and preparation of microwave experiments.

\end{document}